\documentclass[fleqn,10pt]{wlscirep}

 \usepackage{setspace}
\usepackage{multirow}

\usepackage{subcaption}
\usepackage{wrapfig}
\usepackage{mathrsfs}
\usepackage{hyperref}
\usepackage{nameref}
\usepackage[normalem]{ulem}

\usepackage{multicol}

\newcommand{\beginsupplement}{%
        \setcounter{table}{0}
        \renewcommand{\thetable}{\arabic{table}}%
        \setcounter{figure}{0}
        \renewcommand{\thefigure}{\arabic{figure}}%
        \renewcommand{\figurename}{Supplementary Figure} 
        \renewcommand{\tablename}{Supplementary Table} 
     }

\title{
Automatic Identification of Crystal Structures and Interfaces via Artificial-Intelligence-based Electron Microscopy
}

\author[$\dagger$ 1]{Andreas Leitherer\thanks{Contributed equally to this work}}
\newcommand\CoAuthorMark{\footnotemark[\arabic{footnote}]}
\author[2]{Byung Chul Yeo\protect\CoAuthorMark}
\author[3]{Christian H. Liebscher\thanks{Corresponding author}}
\author[1, 4]{Luca M. Ghiringhelli\protect\CoAuthorMark \thanks{Current address: Department of Materials Science and Engineering, Friedrich-Alexander Universität Erlangen-Nürnberg, Germany} }
\affil[1]{NOMAD Laboratory at the Fritz Haber Institute of the Max Planck Society and IRIS-Adlershof of the Humboldt University, Berlin, Germany}
\affil[2]{Department of Energy Resources Engineering, Pukyong National University, Busan 48513, Republic of Korea}
\affil[3]{Max-Planck-Institut für Eisenforschung, Düsseldorf, 40237, Germany}
\affil[4]{Physics Department and IRIS Adlershof, Humboldt-Universität zu Berlin, Berlin, Germany}

\affil[$\dag$]{andreas.leitherer@gmail.com, ghiringhelli@fhi-berlin.mpg.de,liebscher@mpie.de}

 \begin{abstract}
Characterizing crystal structures and interfaces down to the atomic level is an important step for designing advanced materials.
Modern electron microscopy routinely achieves atomic resolution and is capable to resolve complex arrangements of atoms with picometer precision.  
Here, we present AI-STEM, an automatic, artificial-intelligence based method, for accurately identifying key characteristics from atomic-resolution scanning transmission electron microscopy (STEM) images of polycrystalline materials. 
The method is based on a Bayesian convolutional neural network (BNN) that is trained only on simulated images. 
AI-STEM automatically and accurately  identifies crystal structure, lattice orientation, and location of interface regions in synthetic and experimental images. 
The model is trained on cubic and hexagonal crystal structures, yielding classifications and uncertainty estimates, while no explicit information on structural patterns at the interfaces is included during training. 
This work combines principles from probabilistic modeling, deep learning, and information theory, enabling automatic analysis of experimental, atomic-resolution images.
 \end{abstract}
\begin{document}

\flushbottom
\maketitle

\thispagestyle{empty}
\clearpage
\section*{Introduction}
Distinct crystal structures, surfaces, and interfaces in bulk as well as nanomaterials play a key role in tailoring desirable properties in many applications, e.g., catalysis or energy conversion and storage \cite{Harmer2011Science,Zhao2020NatRevMat,Luo2019ACSNano}. 
In particular, exposed surface structures in catalysts determine catalytic performances comprising activity and selectivity \cite{Barroo2020NatCatal}. 
Furthermore, interfaces such as  grain boundaries or stacking faults  can largely affect the transport properties in energy storage or conversion devices\cite{Gordiz2016SciRep,He2021JPhysChemC,Sun2017ACSApplMaterInterfaces,Hu2022EnergyEnvironSci,Lee2018MaterTodayEnergy,Naumann2014SolEnergyMaterSolCells}. 
For example, grain boundaries serve as ion migration paths in batteries \cite{He2021JPhysChemC,Sun2017ACSApplMaterInterfaces}, act as scattering sites for phonons in thermoelectric devices \cite{Gordiz2016SciRep,Hu2022EnergyEnvironSci}, and could degrade electronic conductivity in solar cells \cite{Lee2018MaterTodayEnergy,Naumann2014SolEnergyMaterSolCells}. 
To engineer advanced materials for such applications, it is necessary to characterize their crystalline structure down to the atomic level, including defects or interfaces, local lattice orientations, and distortions \cite{Lu2018AdvMater,Liebscher2018PhysRevLett,Meiners2020Nature}. 
Currently, the ultimate tool to probe  imperfections in crystalline materials is electron microscopy.

To date, electron microscopy techniques with aberration correction have been developed for investigating microstructures of materials with atomic spatial resolution. 
In particular, scanning transmission electron microscopy (STEM) images are more readily interpretable than images obtained via high resolution transmission electron microscopy (HR-TEM), due to direct correlation between image contrast and the atomic number Z of the observed species \cite{Pennycook2011Springer}. 
In STEM, a focused, high-energy electron beam passes through an electron transparent and hence thin sample. 
The electrons interact with the atoms in the sample  and get both scattered elastically and inelastically, enabling to image the sample through various detector geometries (e.g., bright field (BF), dark field (DF), angular dark field (ADF), as well as high-angle annular dark field (HAADF)) and probe it 
through spectroscopic techniques (e.g., electron energy-loss spectroscopy (EELS) and energy-dispersive X-ray spectroscopy (EDS)) \cite{Thomas2015ChemPhysLett,Collins2017Ultramicroscopy,Pan2017JAmChemSoc,Ophus2019MicroscopyandMicroanalysis}. The most commonly employed technique to image atomic structures and crystalline defects is HAADF-STEM, where electrons scattered to large angles are collected by an annular detector forming an incoherent image. 
Moreover, a variety of data channels can be collected simultaneously with high-speed detectors, but as of today the wealth of information available in STEM is not fully exploited, due to the lack of versatile, automatic analysis tools\cite{kalinin2015big, spurgeon2021towards}.

Big-data analytics and artificial-intelligence have great potential for analyzing   
large electron-microscopy data, with several applications to various datasets being reported\cite{Aguiar2019SciAdv,Ziatdinov2019SciAdv,
Vasudevan2018NpjComputMat,Jesse2016SciRep,
Ziatdinov2017ACSNano,spurgeon2021towards,Kalinin2022NatRevMethodsPrimers}. 
Such methods are introduced to uncover overlooked characteristics and this way drive a paradigm shift in image analysis and design of descriptors of atomic-resolution data. 
To provide a few examples, space-group classification was proposed based on electron imaging and diffraction datasets\cite{Aguiar2019SciAdv}. 
Also, multivariate statistical techniques were employed to extract structural information such as the crystal structure and orientation of a small sample region from complex four-dimensional STEM datasets \cite{Jesse2016SciRep}. 
Detection and assignment of microstructural characteristic that differ from the vast majority of crystalline regions and phases in STEM datasets 
has been performed, e.g., the identification of the local dopant distribution in graphene\cite{Ziatdinov2019SciAdv,Ziatdinov2017ACSNano}, or monitoring of electron-beam induced phase transformations\cite{Vasudevan2018NpjComputMat}. 
One can also train AI methods to assign two-dimensional (2D) Bravais lattices to STEM or scanning tunneling microscopy (STM) images\cite{Vasudevan2018NpjComputMat, choudhary2022atomvision}. 
A further approach is to reconstruct the real-space lattice from atomic-resolution images\cite{Ziatdinov2017ACSNano,wei2022benchmark, corrias2022automated}, providing real-space information that can be analyzed with structure-identification methods that are based, for instance, on graphs\cite{Ziatdinov2017ACSNano} or structural descriptors \cite{Leitherer2021NatComm}. 
Unsupervised learning for defect detection or chemical-species classification is reported, for instance, in  \cite{guo2021defect, kalinin2021deep}. 
The above approaches rely heavily on recent developments in deep learning\cite{Goodfellow-et-al-2016}. 
Properly trained neural networks (NNs) such as convolutional neural networks (CNNs) have been shown to solve image classification problems more accurately than other machine-learning methods and in particular, more efficiently than humans, especially in high-throughput tasks.

Here, we propose AI-STEM, which stands for Artificial-Intelligence Scanning Transmission Electron Microscopy. AI-STEM automatically identifies projected crystal symmetry and lattice orientation as well as the location of defects such as grain boundaries in STEM images. 
Both synthetic and experimental images can be processed directly and in automatic fashion, no reconstruction of real-space lattices is required. 
We employ a Fourier-space descriptor, termed FFT-HAADF (FFT: Fast Fourier Transform), as input for a CNN. 
The deep-learning model classifies a given image into a selection of crystalline regions that  differ not only by crystal symmetry but also orientation.  
 This provides additional information compared to, for instance, the classification of a given image into the five Bravais lattices that exist in two dimensions. 
 In particular, we propose an efficient training scheme that enables fast retraining and extension  of the method. 
The model is trained on simulated images only, achieving  near-perfect accuracy on both training and test data (in total $31\,470$ data points, see  Methods). 
The training data contains typical noise sources 
that are encountered in experiment.  
Notably, we adopt the Bayesian neural-network (BNN) approach, employing the Monte Carlo dropout framework that was originally developed by Gal and Ghahramani\cite{Gal2016ICML}. 
BNNs do not only classify a given input but also provide uncertainty estimates. 
We exploit this additional information that is absent in standard deep-learning models to locate bulk regions as areas of low and interfaces as areas of 
high model uncertainty.  
This way, AI-STEM can identify defects without  being explicitly informed about them during training.  
The identification of bulk and interface regions is related to semantic segmentation, a popular computer-vision task in which each image pixel is classified in order to locate individual objects\cite{long2015fully}. 
Based on AI-STEM's bulk-versus-interface segmentation, further analysis can be conducted where it is meaningful \textendash\ according to the model: for instance, we demonstrate how the local lattice rotation can be calculated in the detected bulk regions. 
Finally, we employ unsupervised learning to visualize 
the high-dimensional NN representations in an interpretable, two-dimensional map.  
This reveals that the model separates not only crystalline grains with different symmetry but also different types of interfaces \textendash\ despite never being explicitly instructed to do so. 
All code and data is made publicly available.

\section*{Results}

\subsection*{Development of an automated classification procedure}
\begin{figure*}[h]
 \centering
 \includegraphics[width=\textwidth]{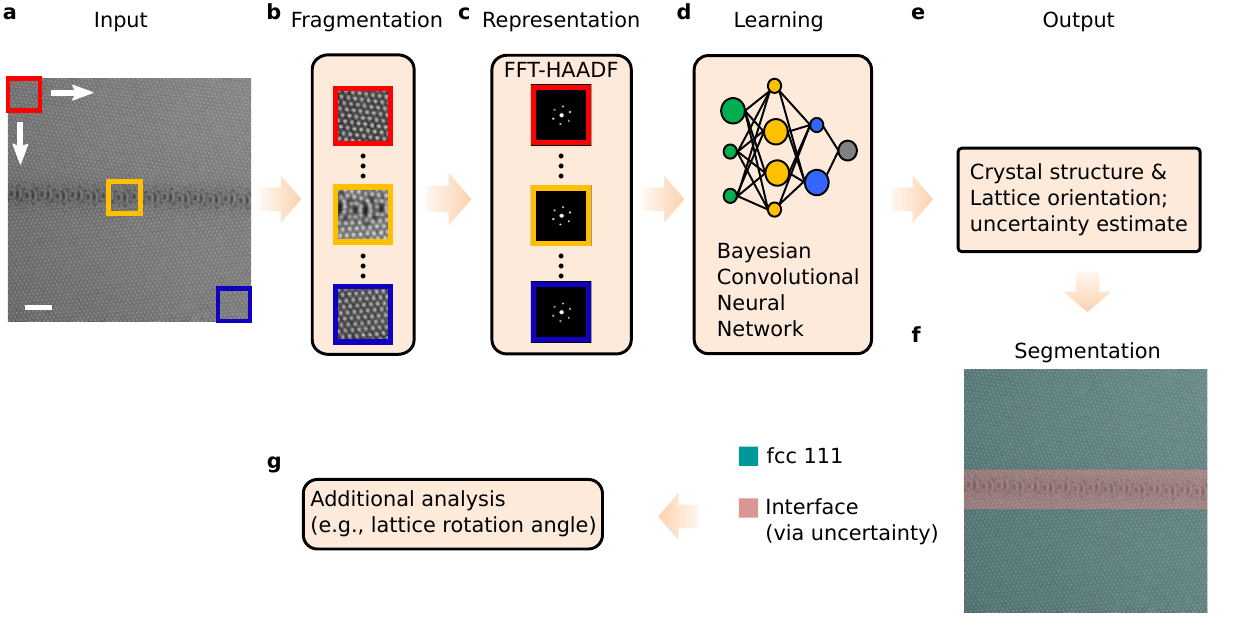}
 \caption{
\textbf{Schematic overview of the AI-STEM procedure for analyzing  experimental STEM images.}
The starting point for AI-STEM (Artificial Intelligence Scanning Transmission Electron Microscopy) is a high-angle annular dark field (HAADF) STEM 
image (\textbf{a}) that here contains two different crystalline regions and one grain boundary (interface). A local window is scanned over the image with a certain stride to fragment the input into local windows (\textbf{b}). Three different local windows are indicated in \textbf{a}, corresponding to regions in the bulk (red, blue) and the boundary (yellow). The local windows are then represented using a fast Fourier transform (FFT) HAADF descriptor (\textbf{c}, normalized between 0 and 1), where typically, a pronounced central peak can be observed. To enhance the neighboring peaks, the  maximum value in the color scale is set to 0.1. 
This Fourier space descriptor is used as input for a Bayesian  convolutional neural network (\textbf{d}) that provides a classification of crystal structure and lattice plane as well as uncertainty estimates (\textbf{e}). The former can be used to detect the bulk regions and the latter reveal the interface and in general, regions with crystal defects (\textbf{f}). On top of this segmentation, additional analysis, e.g. the determination of the local lattice orientation can be performed (\textbf{g}). The scale bar is 1\,nm in \textbf{a}. 
 }
 \label{fig:fig1}
 \end{figure*}

Our goal is to develop an automatic framework for analyzing experimental HAADF-STEM images of bulk materials such as shown in Fig. \ref{fig:fig1}a: 
in this image, the bulk crystalline regions are  separated by a grain boundary (the interface region). 
The final prediction as shown in Fig. \ref{fig:fig1}f  should classify the image into  bulk and interface regions, while also obtaining information about the bulk symmetry and lattice orientation. 
Here, the bulk region should be labeled as ``fcc 111'', i.e., face-centered cubic symmetry in [111] orientation,  since both grains are viewed along their common [111] zone axis corresponding to the tilt axis of the grain boundary. 
Finally, AI-STEM's predictions can be used to automatically identify where to calculate additional properties that provide further characterization, for instance, of the bulk regions and their local lattice rotation (cf. Fig. \ref{fig:fig1}g). 
In the following, we explain the intermediate steps that are required to map image input (Fig. \ref{fig:fig1}a) to a characterization such as shown in Fig. \ref{fig:fig1}f. 

\subsection*{Fourier-space representation of atomic-resolution images}
To achieve sensitivity  to the substructure in an image such as shown in Fig. \ref{fig:fig1}a, we divide it into local fragments (cf. Fig. \ref{fig:fig1}b). 
Specifically, a sliding window of predefined size is scanned over the whole image and local patches are extracted for each stride. 
This allows to investigate structural transitions, e.g., between bulk and interface regions, in a smooth fashion.
The selection of stride and window size is discussed in the Methods section. 
Each of the local patches is then transformed into reciprocal space by computing a Fourier-space descriptor (cf. Fig. \ref{fig:fig1}c). 
Essentially, the fast Fourier transform (FFT) is calculated with additional pre- and post-processing steps (see Methods). 
We term this descriptor FFT-HAADF and use it as input for the machine-learning classification model. 
By calculating the Fourier transform, information on the lattice periodicity is enhanced, thus providing a starting point for a machine-learning model,  which can be generalized to imaging modalities that provide atomic resolution information, such as HR-TEM or STM. 
In addition, translational invariance is introduced already at the level of the representation.  
The descriptor is not rotationally invariant, which is why we employ data augmentation, as we will explain in the section "Training data generation".

\subsection*{The Bayesian classification model}
\begin{figure*}[t]
 \centering
 \includegraphics[width=\textwidth]{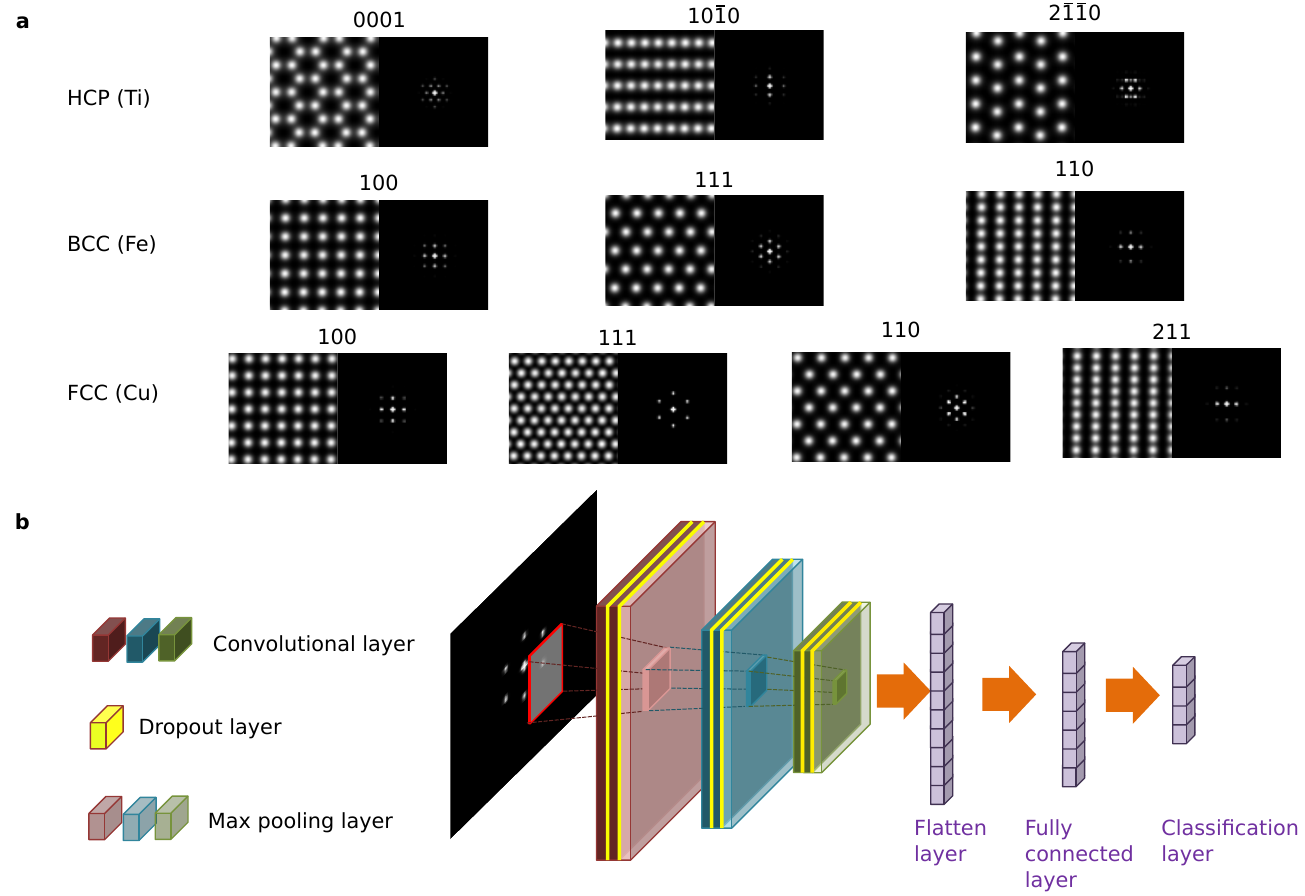}
 \caption{
\textbf{Image descriptor and convolutional neural network (CNN) model  for classification of STEM HAADF images.} 
\textbf{a} Examples of FFT-HAADF images for all 10 crystalline surfaces included in the training set, which include face-centered cubic (fcc), body-centered cubic (bcc) and hexagonal close-packed (hcp) symmetry. \textbf{b} Schematic CNN architecture. FFT-HAADF images are used as the input, and the assignment to one  of the 10 classes is calculated in  the final layer.
 }
 \label{fig:fig2}
 \end{figure*}

To define the classification task,  we need to specify the target labels as well as the model  that maps the FFT-HAADF descriptor to the corresponding target labels. 
As classification model, we employ a CNN. 
This machine-learning method is well-known for its record-breaking performance in image classification\cite{krizhevsky2012imagenet, Lecun2015Nature} and is thus a perfect fit for our problem setting. 
The model receives the FFT-HAADF image descriptor as input and assigns the symmetry (e.g., face-centered cubic) and lattice orientation (e.g., [111], cf. Fig. \ref{fig:fig1}d, e). 
We select in total 10 different crystalline surface structures into which a given image is classified (cf. Fig. \ref{fig:fig2}a). 
This includes the most common crystal structures appearing in metals, comprising face-centered cubic (fcc), body-centered cubic (bcc), and hexagonal close-packed (hcp) structures. 
We focus on low-index crystallographic orientations, which can be resolved at atomic resolution, as the projected interatomic distances are well within the resolution limit. 
The selected orientations are also based on mono-species metal systems for each of the crystal structures considered here: copper (Cu) for fcc, iron (Fe) for bcc, and titanium (Ti) for the hcp structure, respectively. 
The CNN consists of a sequence of convolutional, pooling, and fully connected layers (cf. Fig. \ref{fig:fig2}b). The last layer is composed by 10 neurons, each corresponding to one of the surface classes. 
In particular, the output neurons are normalized such that each represents the classification probability for one of the 10 surface structures. 
For a given image, the most likely class corresponds to the predicted label.
In the complete AI-STEM workflow, the CNN is applied to each local window, providing a classification for each local segment (Fig. \ref{fig:fig1}e). 

In general, beyond classification, it is desirable to estimate the model uncertainty. 
This allows to assess how much one can trust a specific prediction, especially in situations that are different to the training set. 
This can be useful in various scenarios, e.g., for autonomous driving\cite{kendall2016modelling} or medical diagnosis\cite{yang2016fast}. 
In our case, we train the model only on  perfect crystal structures with periodic arrangements of atomic columns and use the uncertainty in the classification to identify the presence of structural defects. 
Given the large number of degrees of freedom for any defect, creating a library of potentially interesting defects for training is challenging \textendash\ which is why we take a different approach:  
 we use a Bayesian neural network\cite{Gal2016ICML, gal2016uncertainty} which does not only classify a given (local) HAADF-STEM image, but also provides uncertainty estimates of the classification.  
If the uncertainty is high (low), the image is likely (unlikely) to 
deviate from the perfect crystal structure (on which the model is trained) and could contain a crystal defect, secondary phase with different crystal symmetry or even amorphous regions.   
This way, we can identify the host crystal structure and orientation at the same time and can locate regions in the image that differ  from any of the training classes, where in this work, we consider grain boundaries as an example.  
One may be tempted to interpret the classification probabilities from the last CNN layer as being informative about model uncertainty. 
However, high classification probability does not always correlate with low uncertainty. In particular, standard NNs are known for overconfident extrapolations \textendash\  even for points that are far outside the training  set\cite{Gal2016ICML, gal2016uncertainty}. 
Modelling of predictive uncertainty can be improved by constructing a probabilistic model that provides a distribution of predictions rather than a single, deterministic one. 

In order to estimate uncertainty in deep learning models, distributions are placed over the NN weights  \textendash\ resulting in probabilistic outputs \textendash\ instead of considering a single set of NN parameters as 
done in the standard approach  \textendash\ resulting in deterministic predictions. 
More formally, a standard NN is a 
 non-linear function  $f_{\boldsymbol\omega}: \mathcal{X} \rightarrow \mathcal{Y}$, i.e., a mapping from input to output space that is parametrized by parameters $\boldsymbol\omega$ 
 (a set of weights and biases 
 $\boldsymbol\omega := \{\mathbf{W}_l, \mathbf{b}_l\}_{l=1}^{L}$, where $L$ is the number of layers).
After training a model on data $D_{\text{train}}$, inference of a  target $y$ (here: a class label) for a new point $\textbf{x}$ (here: the FFT-HAADF image descriptor) is calculated via 

\begin{equation}
p(y| \mathbf{x}, D_{\text{train}}) = \int p(y|\mathbf{x}, \boldsymbol\omega) 
   p(\boldsymbol\omega|D_{\text{train}}) d\boldsymbol\omega.
   \label{eq:1}
   \end{equation}
  In this expression, $p(\boldsymbol\omega|D_{\text{train}})$ denotes the \textit{posterior} that indicates how likely a set of parameters is given training data $D_{\text{train}}$. 
  Moreover, the likelihood $p(y|\mathbf{x}, \boldsymbol\omega)$ corresponds to the \textit{softmax} activation function \textendash\ a standard approach to normalize the output layer such that they can be interpreted as classification probabilities:
  
\begin{equation}
 p(y=c|\mathbf{x}, \boldsymbol\omega) = \frac{\exp{\left( [f_{\boldsymbol\omega}(\mathbf{x})]_c \right)}}{\sum_{c^\prime} \exp{\left( [f_{\boldsymbol\omega}(\mathbf{x})]_{c^\prime} \right)}}.
 \label{eq:2}
\end{equation} 
where $[f_{\boldsymbol\omega}(\mathbf{x})]_c$ is the output value of the NN for the class $c$. 
  We see in Eq. \ref{eq:1} that instead of a single hypothesis, all parameter settings  weighted by their posterior probabilities are included during inference. 
  The standard approach would correspond to choosing the posterior as a delta distribution over a specific parameter setting \textendash\ resulting in the above-mentioned overconfident predictions in out-of-distribution scenarios. 
  Evaluating integrals over the whole parameter space, as appearing in Eq. \ref{eq:1}, is practically impossible \textendash\ especially for large deep learning models. 
  Fortunately, approximating tools for evaluating Eq. \ref{eq:1} are available.
  
  One way to approximate Bayesian inference in deep learning models (i.e., Eq. \ref{eq:1}) is  \textit{Monte Carlo (MC)  dropout}\cite{Gal2016ICML, gal2016uncertainty}. 
  This approach is principled in the sense that the uncertainty estimates from MC dropout approximate those of a Gaussian process\cite{gal2016uncertainty}. 
  In more detail, dropout\cite{hinton2012improving, srivastava2014dropout} is employed \textendash\ a regularization technique that is usually used to avoid overfitting by dropping individual neurons during training. 
  This way, the model has to compensate the loss of individual neurons, avoiding  that the neural activation concentrates to local parts of the network.    
  It has been shown that powerful uncertainty estimates can be obtained by using dropout not only during training but also at test time\cite{Gal2016ICML}. 
  Specifically, for a given input, the output layer is sampled for a certain number of iterations $T$, where each sample is calculated from different networks that are perturbed according to the dropout algorithm. 
  To obtain a Bayesian CNN, dropout is applied after each convolutional and fully connected layer (see the yellow blocks in \ref{fig:fig2}b). 
  Classification can then be performed by calculating a simple average, i.e., the probability of class $c$ given input $\textbf{x}$ and training data $D_{\text{train}}$ (whose general expression is shown in Eq. \ref{eq:1}) can be approximated as  
  
  \begin{equation}
p(y=c|\mathbf{x}, D_{\text{train}}) \approx \frac{1}{T} \sum_{t=1}^{T} p(y=c|\mathbf{x},\boldsymbol\omega_t).
      \label{eq:3}
  \end{equation}
Here, $p(y=c|\mathbf{x},\boldsymbol\omega_t)$ (defined in Eq. \ref{eq:2}) denotes 
the classification probability of class $c$ given input $\mathbf{x}$ and 
parameter configuration $\omega_t$ that is obtained by random removal of neurons (defined according to the dropout algorithm).  
Modest number of samples typically suffice\cite{gal2016uncertainty}, where in this work, we employ $T=100$ samples. 
Notably, this process is in principle trivial to parallelize. We discuss more details on computation time and choice of $T$ in the Supplementary Information (cf. Supplementary Fig. \ref{fig:figS4}). 
Beyond the simple average in Eq. \ref{eq:1},  additional information about the model confidence is contained in the collection of samples $p(y=c|\mathbf{x},\boldsymbol\omega_t)$. 
For this, we invoke information theory, specifically mutual information. 
This (scalar) quantity provides a means to quantify the uncertainty, which has been employed in different settings including self-driving cars\cite{Michelmore2018ArXiv} as well as crystal-structure identification\cite{Leitherer2021NatComm}. 
The mutual information is defined  between predictive and posterior distribution and is denoted as  $I(\boldsymbol\omega, y| D_{\text{train}}, \mathbf{x})$ (see  Methods for exact definition).  
Intuitively, it can be understood as the information gained about the model parameters $\boldsymbol\omega$ if one would receive the label $y$ for a new point $\mathbf{x}$. 
Thus, if the mutual information is high for a given data point, one would gain information once the label is specified \textendash\ corresponding to high predictive uncertainty. 
Similar to Eq. \ref{eq:1}, integrals over the whole parameter space appear, which are computationally intractable. 
However, using MC dropout, one can find a tractable expression that only involves summations over all classes and samples\cite{Gal2016ICML} (Methods).

\subsection*{Training data generation}
To train the  classification model for crystal-structure identification in atomic-resolution images, a  suitable training dataset has to be generated.  
Notably, we refrain from training on experimental images which may contain unknown artefacts, such as noise, distortions or defects. Furthermore, acquiring and curating an experimental database of images of pristine crystal structures imaged at different orientations with atomic resolution is an elaborate task.
 Instead, we train only on simulated images, where we have exact control over imaging conditions and noise sources, allowing us to create a dataset with known labels.  
Obtaining such reliable training data is essential to achieve trustable labeling output of the CNN. 
One may criticize simulations for potentially missing crucial features that are present in experiment. 
However, with the advent of aberration-correction in STEM \cite{Pennycook2011Springer}, the direct comparison of experimental and simulated images at atomic resolution became accessible also on a quantitative basis \cite{LeBeau2008PhysRevLett}. 
It has even been shown that it is possible to determine the number of atoms in an atomic column or to retrieve the 3D atomic structure of nano-objects by combining experimental and simulated images \cite{LeBeau2010NanoLett,Yu2016ACSNano}.
Recently developed efficient implementations of the multislice algorithm enable to simulate STEM images similar to experimental conditions \cite{abtem, Ophus2017AdvStructChemImag}.
Using high-performance computing, realistic simulations of images can be conducted, achieving  computation times of a few hours to days for 10-100 images. 
In this work, we provide additional speed-up by using a convolution-based approach, reducing the computation time from days to minutes for an entire training dataset (see Methods).
Using this efficient simulation scheme, we obtain images for each of the 10 classes for different lattice constants. 
Additionally, we include data augmentation steps to consider a range of lattice rotations and noise sources that are resembling typical experimental conditions (Methods). 
In this work, we include lattice shear, blurring, as well as Gaussian and Poisson noise \textendash\ resulting in $31\,470$ data points. We want to emphasize that even though the model is trained on synthetic data, we apply it to classify experimental atomic resolution STEM images as shown in the Results section (see Fig. \ref{fig:fig4}). 

\subsection*{Neural-network training procedure}
For training, the $31\,470$ data points are split, where 80\% is used for training and $20\%$ for validation. Based on the performance on the validation set, we optimize hyperparameters such as the filter size in the convolutional layers and dropout ratio (the number of neurons dropped). 
Specifically, we employ Bayesian optimization, which is a general approach for global optimization of black-box functions that are computationally expensive to evaluate\cite{ghahramani2015probabilistic}. 
This makes Bayesian optimization a perfect fit for optimizing NNs, where exploring different architectures and optimization parameters is typically accompanied with high computational cost. 
Here, the black-box function to be optimized is the validation loss, and the optimization protocol we invoke\cite{10.5555/3042817.3042832} 
provides us with a list of candidate models, all with near-perfect accuracy (see  Methods). Their uncertainty estimates, however, are different, as we will highlight via the following model selection procedure. 

\begin{figure*}
 \centering
 \includegraphics[width=\textwidth]{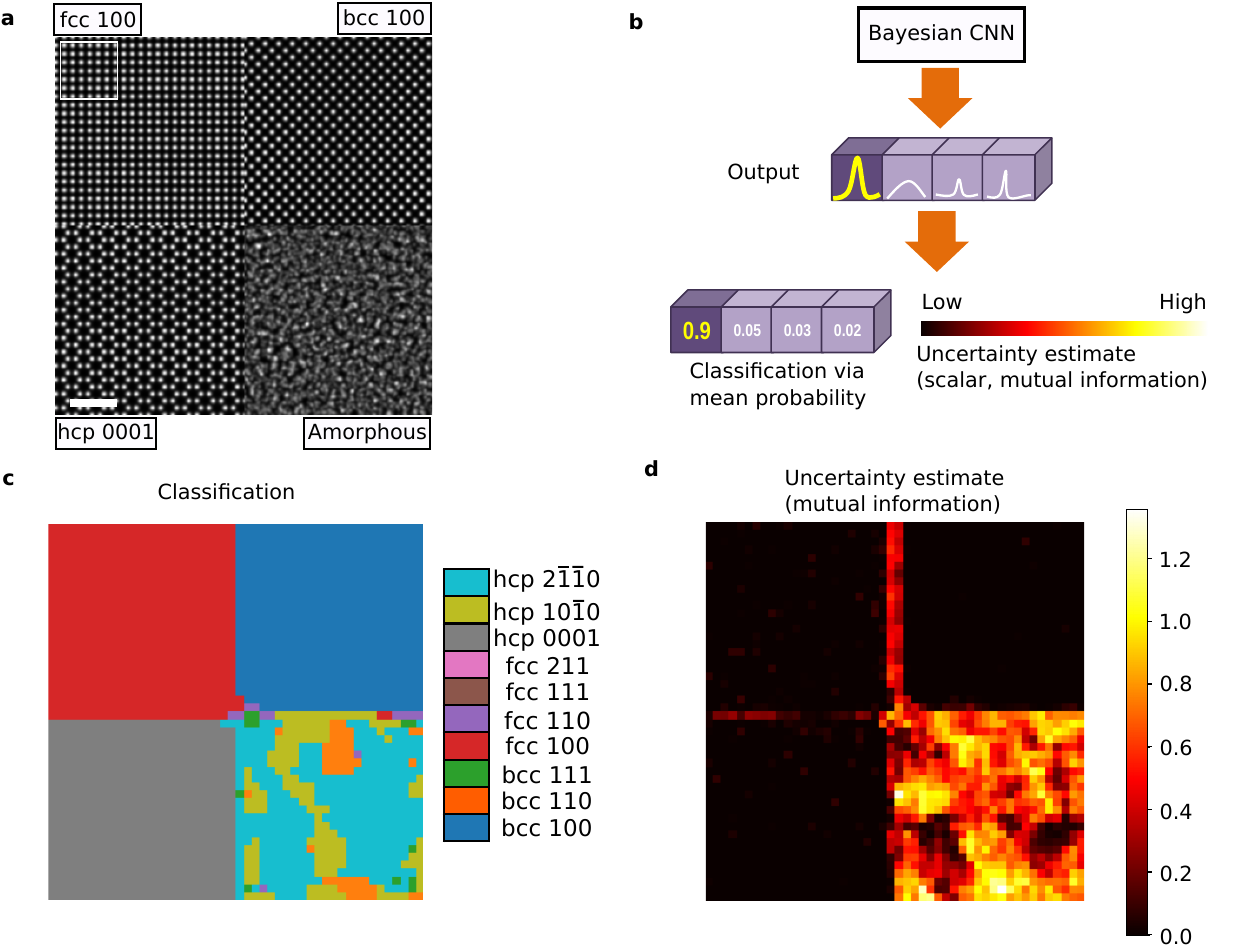}
 \caption{ \textbf{Application of AI-STEM to synthetic, polycrystalline data.}
\textbf{a} The simulated image has 4 crystalline regions with different structural order, including three crystalline (Cu fcc [100], Fe bcc [100], Ti hcp [0001]) and one  amorphous grain. 
Each grain is rectangular with an edge length of 40\,\AA.  
The sliding window is $1.2\times1.2\,\text{nm}$ (100 pixels) and is visualized in the top left corner. 
\textbf{b} The Bayesian CNN employed in the AI-STEM workflow (cf. Fig. \ref{fig:fig1}) provides a distribution and not only point estimates in the final output layer. The averaged classification probabilities can be used to identify the most likely class (\textbf{c}). An uncertainty estimate (a scalar value, cf. \textbf{b} and Eq. \ref{equation:mutual_information}) can be obtained via the mutual information (\textbf{d}), revealing the grain boundaries as well as the amorphous region.
The scale bar is 1\,nm in \textbf{a}. 
}
 \label{fig:fig3}
 \end{figure*}

To find the model that shows strongest performance in both classification and detection of out-of-training-distribution regions, we analyze the simulated test image in Fig. \ref{fig:fig3}a. 
It contains both crystalline and amorphous regions, providing a test bed for identifying models with high uncertainty at the transition between grains and in the amorphous region \textendash\ both of which are  never shown to the models during training.  
The four regions in the image are simulated separately (using full multi-slice simulations) and then stitched together.  
Three of these regions are crystalline, representing one of the in total three different symmetries in the training set: Fe (bcc, [100]), Cu (fcc, [100]), and Ti (hcp, [0001]). 
Here, we expect low uncertainty and correct assignment of the respective symmetry. 
The amorphous region is simulated based on a three-dimensional structure obtained via  realistic molecular-dynamics simulations of amorphous silicon\cite{deringer2018realistic}.
All models obtained via Bayesian optimization are applied to this image. 
Given their near-perfect accuracy during training, they all can recognize the crystalline parts of the image, while their assignments in the amorphous region differ. 
We can now also analyze the corresponding uncertainties, which provide an estimate of the reliability of the classifications  
  (cf. Fig. \ref{fig:fig3}b). 
We select the model with the highest uncertainty, as quantified by the mutual information (cf. Eq. \ref{equation:mutual_information}), in the amorphous region. 
For this model, the classification results are shown in Fig. \ref{fig:fig3}c, where one can see that the correct crystal symmetries are assigned in the expected regions, while in the amorphous part, several different phases are assigned. 
The mutual information shown in Fig. \ref{fig:fig3}d increases at the interfaces between the four different crystalline regions, as well as in the amorphous part. 
The detailed architecture is specified  in  Table \ref{table:ai4stem_cnn}.

\subsection*{Application to experimental STEM data}

Now we turn to applying AI-STEM to experimental data.  
In the following, we challenge the model with several HAADF-STEM images, demonstrating the practical applicability of AI-STEM. In particular, we show that the model can classify crystalline regions in experimental images and  how the bulk-versus-interface segmentation can be inferred and employed for further analysis \textendash\ here, for determining the local lattice orientation in the bulk regions. 

First, a HAADF image of elemental Cu shown in Fig. \ref{fig:fig4}a is analyzed. 
The image contains a horizontally aligned grain boundary separating two misoriented single crystals with a [111] orientation in the upper and lower grain, respectively. 
As shown in Fig. \ref{fig:fig4}b, the model classifies the grain regions correctly as fcc [111].
At the interface, the same label is assigned, but 
with increased uncertainty (as quantified by  mutual information), allowing to detect the interface region (cf. Fig \ref{fig:fig4}c). 

The segmentation obtained via AI-STEM's predictions can now be used to conduct further analysis of the local lattice structure. 
Practically, to separate the image into bulk and interface regions, we fix a mutual-information threshold of 0.1, interpreting all local windows above this value as interface and the remaining ones as bulk. 
Depending on the type of region, i.e., interface or bulk, different quantities are suited. 
As an example, we calculate here the local lattice orientation, a quantity that is only reasonable to compute in the bulk regions. 
Specifically, for each local window, we reconstruct\cite{nord2017atomap} the real-space lattice from the atomic columns and determine\cite{myronenko2010point, gatti2022pycpd}  the angle of misalignment with respect to a reference training image or rather its reconstructed atomic columns (cf. Supplementary Methods for more details).  
Note that this way, information from the training data is entering this analysis. 
Also note that the reference lattice is not required as input but determined based on the NN assignments \textendash\ making this procedure fully automatic and extendable (in case of retraining and new classes being added to the training set).
The calculated angle is termed lattice mismatch and  the results for the Cu grain boundary are shown in Fig. \ref{fig:fig4}d. 
The reference images are shown below the heatmaps. 
For the interface region, depicted in gray in Fig. \ref{fig:fig4}d, no calculation is performed. 
The expected misorientations are exemplarily indicated in Fig. \ref{fig:fig4}a, which closely match the calculated values of Fig. \ref{fig:fig4}d.

Next, we consider a HAADF image of Fe\cite{Ahmadian2018NatComm} containing a grain boundary that is horizontally aligned and separates two crystalline grains with [100] orientation (cf. Fig. \ref{fig:fig4}e).  
Compared to the previous example, this image contains   intensity variation of the background but also the atomic columns, which is more pronounced, for instance, in the upper left part compared to the lower part of the image. Such variations are common in experimental images and may stem from surface damage induced during sample preparation or surface oxide formation. 
However, AI-STEM correctly classifies the bulk regions as bcc [100] (cf. Fig. \ref{fig:fig4}f), while the assignment changes at the grain boundary, but with increased uncertainty (cf. Fig. \ref{fig:fig4}g). 
In the upper left in more noisy parts of the image, the uncertainty also increases. 
The obtained mismatch angles 
are shown in Fig. \ref{fig:fig4}h and also here calculated and expected angles (again indicated in the original image in Fig. \ref{fig:fig4}e) are in agreement.  

Finally, we investigate a low-angle [0001] tilt grain boundary in Ti (cf. \ref{fig:fig4}i), which consists of a periodic array of dislocations with a line direction perpendicular to [0001]. 
Hence, the interface structure is qualitatively different compared to the previously shown high angle grain boundaries for Cu and Fe. 
In particular, the smaller misorientation angle between both grains in the Ti image leads to regions within the interface where the atomic lattices of the two grains are still connected with each other.
AI-STEM correctly assigns hcp [0001] (cf. Fig. \ref{fig:fig4}j), with only few outliers in the classification at the grain boundary, which is again revealed via the mutual information (cf. Fig. \ref{fig:fig4}k). 
One can observe that the mutual information is decreasing in the regions in between the grain boundary dislocations, where the lattice resembles that of undisturbed Ti [0001] and is increasing in the locations of the dislocation cores at the interface. 
This shows that the uncertainty estimate of the predictions can even be used to locate more confined lattice defects such as individual dislocations. Similar to the previous two examples, we obtain the local lattice mismatch (cf. Fig. \ref{fig:fig4}l) that matches the expectations (cf. Fig. \ref{fig:fig4}i) with a margin of few degrees.

\subsection*{Analyzing AI-STEM's internal representations via unsupervised analysis}
So far, we have demonstrated how AI-STEM can be used to classify lattice symmetry and orientation and is capable of detecting interfaces and even individual dislocations within an interface. 
To understand how the model interprets crystalline grains and interface regions, we apply unsupervised learning to the internal NN representations. 
Specifically, we employ manifold learning to embed the high-dimensional NN representations into two-dimensional, readily interpretable maps. 
We employ Uniform Manifold Approximation and Projection (UMAP)\cite{mcinnes2018umap}, which approximates the manifold that underlies a given dataset, and allows to construct low-dimensional embeddings that can capture both global and local relationships among the original, high-dimensional data points. 
We consider the experimental images shown in Fig. \ref{fig:fig4}a, e, i, and compute the NN representations for each of the local windows, as determined within the AI-STEM workflow (cf. Fig. \ref{fig:fig1}b). 
Superficially, we inspect the last, fully connected layer before the output layer, i.e., before the classification is conducted (cf. Fig. \ref{fig:fig2}b). 
The two-dimensional UMAP embedding is shown in Fig. \ref{fig:fig5}a, where the color scale corresponds to the NN assignments. 
Despite the high level of compression, from 128 to 2 dimensions, all three images are well separated.
For each image, two sub-clusters can be observed that correspond to the two bulk grains (cf. Fig. \ref{fig:fig5}a). 
These are joined by contiguous strings that correspond to the interface regions, respectively. 
This is also visualized by using the mutual information as a color scale (cf. Fig. \ref{fig:fig5}b), where along the strings, increased uncertainty can be observed (which indicates the presence of the defects). 
Notably, the different grain boundary types (e.g. high angle vs. low angle) are also mapped to different regions in the map.
This demonstrates the capability of AI-STEM to not only recognize bulk symmetry and orientation but also to distinguish different interface types \textendash\ even though it has never been provided with explicit examples for such a task during training.

\section*{Discussion}
In this work, we propose AI-STEM which  automatically characterizes crystal structure 
and interfaces in simulated and experimental atomic-resolution STEM datasets. 
This is enabled by adapting several techniques: we employ signal-processing tools to represent imaging data, deep learning to identify crystal symmetry and orientation, and Bayesian modeling in combination with information theory to estimate model uncertainty as well as to optimize NN hyperparameters. 
At the core of AI-STEM is a Bayesian convolutional neural network, which goes beyond standard NN models, providing classifications and principled uncertainty estimates.  
The former allow identification of lattice symmetry and crystal orientation while the latter are used to segment an image into bulk and interface regions. 
Despite being trained only on simulated STEM images of perfect lattice structures, AI-STEM generalizes to experimental images, as demonstrated by several challenging examples. 
 The training data can be obtained by discrete multislice image simulations considering dynamical scattering effects, while, in this work, we show that a fast convolution approach can be employed.  
In order to verify the applicability of the labeling procedure, a diverse set of simulated images  of  typical monocrystalline structures is generated, serving as reliable ground truth. 
Based on the segmentation provided by AI-STEM's prediction, one can conduct augmenting  analysis that reveals additional characteristics of the identified regions. 
Here, we determine the local lattice rotation in the crystalline grains. 
Using unsupervised learning, we demonstrate that different types of interfaces appear separated in the internal NN space, despite no explicit information on any interface pattern is being provided during training.  
This analysis also shows how unsupervised learning can be used to \textit{explain} a black-box model, in post-hoc fashion\cite{lipton2018mythos, murdoch2019definitions, roscher2020explainable}. 
Moreover, on-line data processing is feasible with the proposed approach since the method is easy to parallelize and already using a single GPU we are within the range of typical acquisition times (cf. Supplementary Fig. \ref{fig:figS4}).

Furthermore, note that the presented experimental images in Fig. \ref{fig:fig4} have near to perfect zone axis orientation within the experimental limit, since the aim is to resolve the atomic structure of the interfaces with highest possible precision. 
While this might be considered an idealized scenario, the results of Fig. \ref{fig:fig4} should at least constitute a test for small
deviations in crystal tilts, which are typically present in experimental images. Since we did not include any information on this in the training data set, the model already at that level
shows to be robust. 
We have conducted further tests for larger deviations in tilt of the adjoining crystals in Supplementary Figure \ref{fig:figS5}. The model provides the expected assignments, even if only lattice fringes are resolved, while the accompanying high uncertainty values require a careful interpretation of the prediction. The model robustness may be further improved by including crystal tilt variations as additional parameters in the training set.

Since various types of noise components, such as scan and detector read-out noise, are typically present in STEM images, we further tested the applicability of AI-STEM for different noise levels as shown in Supplementary Figure \ref{fig:figS6}. Here, we deployed AI-STEM to images with different degrees of primarily fast scan noise that are contained in the images. 
Specifically, we consider the experimental image in Fig. \ref{fig:fig4}a as a reference example with reduced fast scan noise by frame averaging and show that the number of frame averages does not influence AI-STEM's performance. Even a single shot frame with high noise contributions is correctly classified showing low uncertainty values in the prediction of the bulk crystal regions (see Supplementary Figure \ref{fig:figS6}).

In the future, it would be interesting to provide not only a bulk-versus-interface segmentation but also predict additional details automatically, e.g., how the crystalline grains differ. 
Currently, this can only be done by additional analysis, e.g., based on reconstructing the (projected) real-space lattice.
However, one can see from the latent space visualization in Fig. \ref{fig:fig5} 
that grains with different orientation are separated. 
In principle, a clustering algorithm may be employed to separate the grains, while this can be challenging to automate as clustering typically involves several parameter choices that are not guaranteed to generalize well. 
Alternatively, one may consider a multi-label classification problem or construct a separate machine-learning model to predict the (local) lattice rotation automatically.  

In conclusion, our method shows great potential to automatically analyze and classify crystallographic attributes in STEM datasets without human intervention. In electron-microscopy research, the development of a “self-driving” microscope appears on the horizon due to rapid advances in artificial intelligence\cite{Dyck2019MRSBulletin,Kalinin2016Nature}. While we focus on mono-species systems as a proof of concept, this work paves the way to autonomous investigations of 
complex nanostructures at the atomic level.

\section*{Methods}

\textbf{AI-STEM parameters}
Besides the classification model, the two most important components are the stride and box size.
For the box size, we recommend a value of 12\,\AA, on which the model is trained. 
If significantly larger window sizes are necessary for a desired application, the practical approach is to augment the dataset using our efficient  training procedure and retrain the model. 
Also note that the model is trained for a specific resolution, in which 1 pixel corresponds to 0.12\,\AA.
For different resolutions, one may simply rescale the image or, as we proceeded here, adjust the window size. 
For instance, the Cu image in Fig. \ref{fig:fig4}a is measured for a resolution of about 0.0880\,\AA per pixel, while the other images in Fig. \ref{fig:fig4} are measured for 0.1245\,\AA. 
To match both resolutions to the training range, we decrease the box size to 136 pixels for Cu (as it is measured at higher resolution, i.e., we need to increase the box size to obtain a number of atomic columns that is comparable to the training set), and 96 for the other two images (as it is recorded at lower resolution, i.e., we smaller windows are required to obtain a number of atomic columns that is comparable to the training set).  
In principle, our data-generation method also allows to vary the resolution, such that retraining with various resolutions could be done as well.
For the stride, we use values on the order of 1\,\AA, to demonstrate the high-resolution capabilities of the approach.
Smaller strides can suffice to reveal the main characteristics, cf. Supplementary Fig. \ref{fig:figS3} (in particular, it is possible to separate an image into bulk and interface regions). 
For the synthetic image in Fig. \ref{fig:fig3}, we employ a stride of 12 pixels, corresponding to $\sim 1.4\,\AA$. 
The same settings were used for the experimental images of Ti and Cu (Fig. \ref{fig:fig4}a, c). For Fe (Fig. \ref{fig:fig4}b), the stride was halved
as this image is smaller (about half of the size of Ti, and two third of Cu), enabling comparable number of local fragments.

\textbf{FFT-HAADF descriptor}
We start from the periodic arrangement of atomic columns in HAADF-STEM
images. 
These are acquired in low-index crystallographic orientations, which directly represent the underlying projected crystal symmetry. 
In the AI-STEM workflow, an input image corresponds to a local fragment or window, extracted from a larger image. 
The cutting procedure may lead to to boundary effects, e.g., truncated atomic columns. This can lead to spurious patterns in the FFT, which is why we apply a window function to the STEM HAADF image before calculating the FFT \textendash\ a standard practice in signal processing\cite{harris1978use}. 
Here, we use the Hann window that provides a smooth decay at the image boundaries. Then, the FFT is calculated, resulting in spectra which have a dominant central peak, suppressing possibly valuable information at higher frequencies. 
Thus, we apply a thresholding scheme: the FFTs are normalized to the range [0, 1] and then all values above 0.1 are set to 1.0.
This provides visible enhancement of peak patterns around the central peak, which is visualized for all classes in this work in  Supplementary Fig. \ref{fig:figS1}.
  
\textbf{Neural network training}
The CNN is trained on $31\,470$ $64\times64$ pixel images (the FFT-HAADF descriptor of the STEM HAADF images). 
A split of this dataset into training and test is performed in stratified fashion (via scikit-learn, using a random state of 42; see Data availability for the dataset link). 
Adam optimization is employed for training\cite{kingma2014adam}. 
The CNN is implemented using Tensorflow\cite{abadi2016tensorflow}. 
 Hyperparameters are optimized using Bayesian optimization, specifically the 
Tree-structured Parzen estimator (TPE) algorithm 
as provided by the python library hyperopt\cite{10.5555/3042817.3042832}. 
We experimented with minimizing either validation loss or accuracy, while no significant difference could be found, in terms of classification accuracy. 
We chose the validation loss as objective function to be minimized.  
We tested different configuration spaces for the network architecture and optimization parameters, including 
number of layers, number of filters, filter size, dropout ratio as well as batch sizes (example notebooks are provided, cf. ``Data availability''). 
The models typically converge to near-perfect accuracy in few epochs and we find that we can restrict to smaller configuration spaces, reducing the computational cost. 
We fix the architecture to 6 layers (number of filters: 32, 32, 16, 16, 8, 8) and focus on the search for the right kernel size ( $3\times3, 5\times5, 7\times7$) as well as the dropout ratio (values between 2 and 10 percent, step size 1 percent). 
In particular, the choice of dropout ratio is known to be important for the quality of the uncertainty estimates\cite{gal2016uncertainty}.
We run the TPE algorithm for 25 iterations. 
Each model is optimized for 25 epochs, saving only the model with best validation accuracy.  
These models achieve all near-perfect accuracy (99,9\,\% classification accuracy on both training and validation set), but their uncertainty estimates differ. 
We thus select the model that has the highest median uncertainty in the amorphous region in the synthetic polycrystal example (Fig. \ref{fig:fig3}), where we expect a low degree of crystallinity. 
The model chosen in this fashion is reported in Table \ref{table:ai4stem_cnn}.

\textbf{Uncertainty quantification}
Given the test point $\mathbf{x}$, the mutual information between the predictions and the model posterior $p(\boldsymbol\omega|D_{\text{train}})$ 
is defined as\cite{houlsby2011bayesian, Gal2016ICML, gal2016uncertainty} 

\begin{equation}
 \mathbb{I} \left[y, \boldsymbol\omega \vert \mathbf{x}, D_{\text{train}}\right] := 
 \mathbb{H}[y|\mathbf{x}, D_{\text{train}}] - \mathbb{E}_{p(\boldsymbol\omega|D_{\text{train}})}\left[ \mathbb{H}[y|\mathbf{x}, \boldsymbol\omega] \right].
 \label{equation:mutual_information_origdef}
\end{equation}
The first term on the r.h.s. is termed \textit{predictive entropy}\cite{gal2016uncertainty}. 
It quantifies  the (average) information in the distribution of predictions and is defined by  

\begin{equation}
 \mathbb{H}[y|\mathbf{x}, D_{\text{train}}] := - \sum_c p(y=c|\mathbf{x}, D_{\text{train}}) \log{p(y=c|\mathbf{x}, D_{\text{train}}}). \label{equation:predictive_entropy}
\end{equation}
The second term on the r.h.s. of Eq. \ref{equation:mutual_information_origdef} is defined as

\begin{equation*}
\begin{aligned}
 & \mathbb{E}_{p(\boldsymbol\omega|D_{\text{train}})}\left[ \mathbb{H}[y|\mathbf{x}, \boldsymbol\omega] \right] :=  \\
& \mathbb{E}_{p(\boldsymbol\omega|D_\text{train})}\left[ \sum_c p(y=c|\mathbf{x}, \boldsymbol\omega) \log  p(y=c|\mathbf{x}, \boldsymbol\omega) \right].
\end{aligned}
\end{equation*}
One may refer to this as expected entropy as it 
averages the entropy of the predictions given 
the parameters $\boldsymbol\omega$ that are distributed according to the posterior distribution\cite{smith2018understanding}. 
Using Monte Carlo dropout, one can approximate the mutual information as\cite{Gal2016ICML} 

\begin{equation}
\begin{aligned}
& \mathbb{I} \left[y, \boldsymbol\omega \vert \mathbf{x}, D_{\text{train}}\right] \approx  \\ 
& - \sum_{c} \left( \dfrac{1}{T} \sum_{t} p \left(y=c \vert \mathbf{x}, \boldsymbol{\boldsymbol\omega}_t \right) \right)\log  \left( \dfrac{1}{T} \sum_{t} p \left(y=c \vert \mathbf{x}, \boldsymbol{\boldsymbol\omega}_t \right) \right) \\
& + \frac{1}{T} \sum_{c}\sum_{t} p \left(y=c \vert \mathbf{x}, {\boldsymbol\omega}_t \right) \log p \left(y=c \vert \mathbf{x}, {\boldsymbol\omega}_t \right). 
\label{equation:mutual_information}
\end{aligned}
\end{equation}

\textbf{Details on training data generation}
For each of the 10 surface classes, we consider a small interval of $\pm0.1\,\text{\AA}$ around their respective experimental lattice parameters. 
This is due to the fact that some of the classes can be similar (a consequence of the 2D projection provided by STEM images), for instance fcc100 and bcc100 (cf. Fig. \ref{fig:fig2}a). 
The lattice parameters are the following: 
for all Cu fcc single crystals, the lattice constant $a$ is 3.63 Å; for all Fe bcc single crystals, the lattice constant $a$ is 2.87 Å; for all Ti hcp single crystals, the lattice constants $a$ and $c$ are 2.95 Å and 4.68 Å, respectively ($c/a\sim 1.587$). 
For each of these classes, we include a range of rotations (0-90 degrees, step size 5 degrees, using the Python package scipy\cite{2020SciPy-NMeth}). 
Then, different noise sources are applied, as implemented in the Python package scikit-image\cite{van2014scikit}: first, shear is applied to all images (affine transformation  applied to the images, only using shear but no scaling or translation) for all rotations. 
We apply additional noise sources for a subselection of data points (only every second rotated and sheared image, keeping the dataset size below 100k), including Gaussian blurring (scanning a Gaussian filter of certain width over the image), and finally, addition of random noise sources (Gaussian or Poisson). Visual examples are provided in Supplementary Fig. \ref{fig:figS2}.

\textbf{Simulation of STEM dataset}
To generate the ground truth STEM datasets consisting of HAADF images, STEM image simulations were performed with the abTEM software package \cite{abtem}. The crystal orientations as shown in Fig. \ref{fig:fig2} for the ten classes are generated by the atomic simulation environment (ASE) python module \cite{Larsen2017JPhysCondensMatter}. The thickness of all simulation cells was set to 8 nm ($z-$direction) with a slice thickness of 0.2 nm. The $x$- and $y-$dimensions of the simulation cells was chosen to be $\sim$ 8 nm, respectively. An electron energy of 300 kV, a probe semi-convergence angle of 24 mrad and semi-collection angles of the HAADF detector ranging from 78 to 200 mrad were used for the simulations. The pixel size was fixed to 12 pm resulting in images with $\sim$ 64 $\times$ 64 pixels. Thermal diffuse scattering was considered by using 12 frozen phonon configurations with a root-mean-squared thermal displacement according to the Debye-Waller factors obtained from Peng et al. \cite{peng1996debye} for Cu, Fe and Ti at 280 K. The image simulations were performed on a Windows 11 Pro based workstation with an Intel Xeon CPU with 32 GB of RAM and a NVIDIA Quadro K1200 GPU. The total simulation times for the Cu-fcc class was $\sim$ 11 hours, for the Fe-bcc class $\sim$ 26 hours and Ti-hcp class $\sim$ 13 hours, respectively.

To speed up the training dataset generation, we also employed a simple convolution approach where the probe wave function generated in abTEM\cite{abtem} was convolved with the summed projected potentials for each cell. This reduces the total calculation time for each class to several minutes. 
We employ this approach for training the CNN model, demonstrating that this computationally efficient approach can yield strong performance on experimental images (cf. Fig. \ref{fig:fig4}).

\noindent\textbf{Scanning transmission electron microscopy (STEM) experiment}
All experimental STEM data were acquired using a probe corrected Titan Themis 60-300 (Thermo Fisher Scientific). 
The TEM is equipped with a high brightness field emission gun and a gun monochromator. The electrons were accelerated to 300 kV and images were recorded at a probe current of ~80 pA with a high-angle annular dark field (HAADF) detector (Fishione Instruments Model 3000). 
The collection angles for the HAADF images were set to 73–200 mrad using a semi-convergence angles of 17 mrad and 23.8 mrad. 
Image series with 20–40 images and a dwell time of 1–2 $\mu$s were acquired, registered and averaged in order to minimize the effect of instrumental instabilities and noise in the images.

Experimental HAADF-STEM images of a $\Sigma\,19b$ $(178)$ $[111]$ tilt grain boundary in Cu with a misorientation angle of $\sim48^\circ\,$, a $\Sigma\,5$ $(013)$ $[001]$ tilt boundary in Fe with a misorientation angle of $\sim38^\circ\,$
and a low angle [0001] tilt GB in Ti with a misorientation angle of $\sim13^\circ\,$ are used to test the ai4stem approach. Details on the sample fabrication and preparation for the Cu, Fe and Ti grain boundary images can be found in \cite{Meiners2020Nature, Ahmadian2018NatComm, devulapalli2021ti}, respectively.

\section*{Data availability statement}
All relevant data (training and test set, experimental and synthetic images, as well as neural-network models) are available at \url{ https://doi.org/10.5281/zenodo.7756516}.

\section*{Code availability statement}
Code and several examples for applying ai4stem and reproducing the results of this article are available at \url{https://github.com/AndreasLeitherer/ai4stem}.

\noindent

\section*{Acknowledgements}
We acknowledge and thank Matthias Scheffler, Angelo Ziletti, Niels Cautaerts, Donghun Kim, and Christoph Freysoldt for inspiring discussions and suggestions.
We acknowledge funding from BiGmax, the Max Planck Society's Research Network on Big-Data-Driven Materials Science. L.M.G. acknowledges funding from the European Union's Horizon 2020 research and innovation program, under grant agreements No. 951786 (NOMAD CoE) and No. 740233 (TEC1p).
Furthermore, the authors acknowledge the Max Planck Computing and Data facility (MPCDF) for 
computational resources and support, which enabled neural-network training    
on 1 GPU (Tesla Volta V100 32GB) on the Talos machine learning cluster. 
B.C.Y. acknowledges funding from the National Research Foundation (NRF) of Korea under Project Number 2021M3A7C2090586.

\section*{Author contributions}

A. L. and B.C.Y. contributed equally to this work.
A.L., B.C.Y., C.H.L., and L.M.G. designed the idea and the project. A.L. and B.C.Y. performed the neural-network calculations. C.H.L. conducted the multi-slice image simulations. 
C.H.L. and L.M.G. supervised the project.
All authors contributed to the manuscript.

\section*{Competing Interests}
The authors declare no competing financial or non-financial interests.

\section*{Figure captions}

\begin{figure*}[!t]
 \centering
 \includegraphics[width=\textwidth]{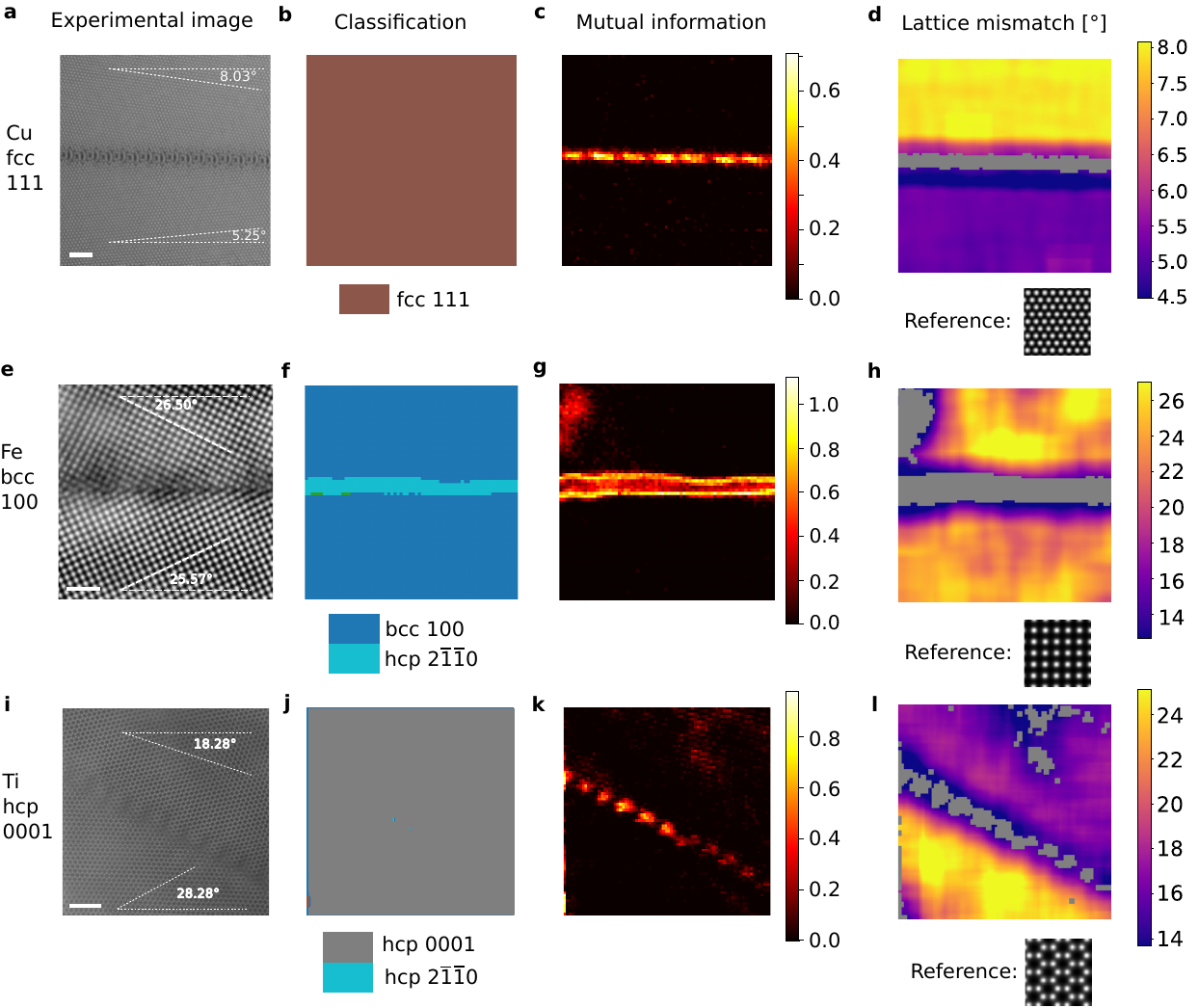}
 \caption{
\textbf{Application of AI-STEM to experimental STEM images with grain boundaries. }
Three experimental images are investigated: 
a $\Sigma\,19b$ $(178)$ $[111]$ tilt grain boundary (GB) in Cu with a misorientation angle of $\sim48^\circ\,$ (class: fcc 111, $9\times9\,\text{nm}$), 
a $\Sigma\,5$ $(013)$ $[001]$ tilt GB in Fe with misorientation angle of $\sim38^\circ\,$ (class: bcc 100, $6.4\times6.4\,\text{nm}$),
and a low angle [0001] tilt GB in Ti with a misorientation angle of $\sim13^\circ\,$ (class: hcp 0001, $12.8\times12.8\,\text{nm}$), which are shown in 
 \textbf{a}, \textbf{e}, and \textbf{i}.  
One can see from the classification maps (\textbf{b}, \textbf{f}, \textbf{j}) that the expected bulk symmetries are
correctly assigned. In the color scale, only the most frequent assignments are labeled, the full color scale (indicating the other assignments, e.g., in \textbf{f} at the interface) is shown in Fig. \ref{fig:fig3}c. 
The uncertainty, as quantified by the mutual information (\textbf{c}, \textbf{g}, \textbf{k}), indicates the grain-boundary regions. Combining these two pieces of information allows to identify the bulk and boundary regions. For the bulk regions, one can conduct further analysis:  as an example, in \textbf{d}, \textbf{h}, \textbf{l}, we determine for each local window the local lattice mismatch, which is defined as the mismatch angle between the real-space lattices 
reconstructed from local window and reference image (shown below the heatmaps in \textbf{d}, \textbf{h}, \textbf{l}). 
This analysis is only conducted where it is meaningful, i.e., in bulk regions,  in particular excluding high-uncertainty regions that are indicated by gray areas in \textbf{d}, \textbf{h}, \textbf{l}. 
The scale bar is 1\,nm in \textbf{a}, \textbf{e} and 2\,nm in \textbf{i}. 
}
 \label{fig:fig4}
 \end{figure*}

 \begin{figure*}[!t]
 \centering
 \includegraphics[width=\textwidth]{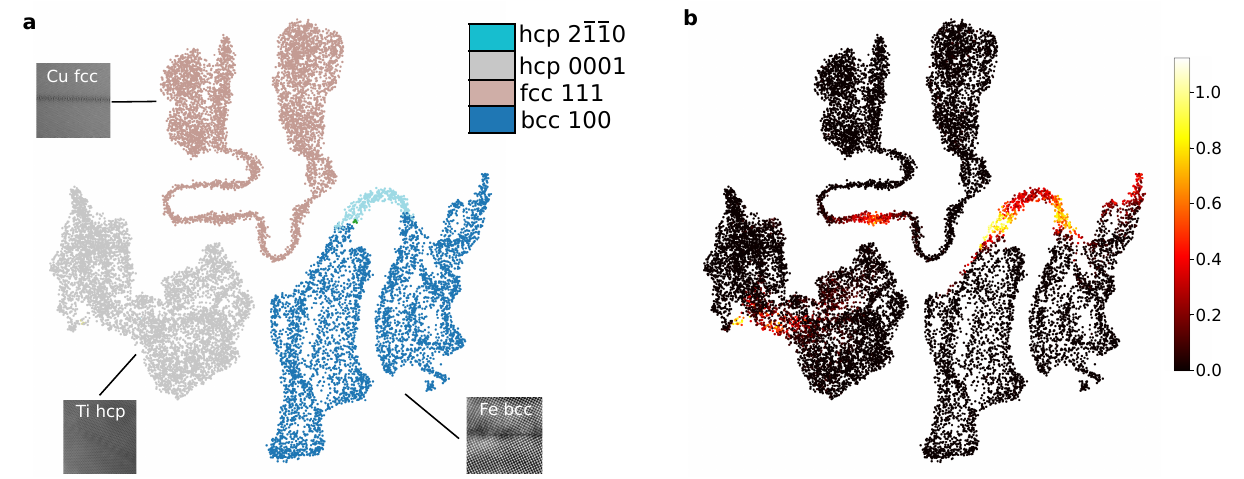}
 \caption{\textbf{Visualizing neural-network representations of local crystalline and defective atomic structure in experimental images.} For each of the three experimental images in Fig. \ref{fig:fig4}, we 
 apply the fragmentation procedure of AI-STEM (cf. Fig. \ref{fig:fig1}b), and extract the neural-network representations of these local windows (for the last fully connected layer before the classification, cf. Fig. \ref{fig:fig2}b). The dimension-reduction (via Uniform Manifold Approximation and Projection, short UMAP) of these high-dimensional NN representation is shown in 
 Fig. \textbf{a} and \textbf{b}, where in \textbf{a} the color scale corresponds to the AI-STEM assignments, and in \textbf{b} the color scale corresponds to the mutual information that quantifies model uncertainty. All images are separated into three connected regions. In each of these, two connected clusters can be seen that correspond to the crystalline grains, while the connections indicate the grain boundary region. Notably, the boundary regions, which correspond to distinct interface types and are of critical importance for the material properties, do not intersect and are thus not confused by AI-STEM.  
 }
 \label{fig:fig5}
 \end{figure*}

\clearpage

\begin{table*}[h]
\centering
\begin{tabular}{ll}
\hline
Layer type & Specifications \\
\hline
\hline
Convolutional layer & 32 filters, $3\times3$ kernel size, $1\times1$ stride, ReLU activation, dropout\\
Convolutional layer & 32 filters, $3\times3$ kernel size, $1\times1$ stride, ReLU activation, dropout\\
Max pooling layer  & $2\times2$ pool size, $2\times2$ stride\\
Convolutional layer & 16 filters, $3\times3$ kernel size, $1\times1$ stride, ReLU activation, dropout\\
Convolutional layer &16 filters, $3\times3$ kernel size, $1\times1$ stride, ReLU activation, dropout\\
Max pooling layer  & $2\times2$ pool size, $2\times2$ stride\\
Convolutional layer & 8 filters, $3\times3$ kernel size, $1\times1$ stride, ReLU activation, dropout\\
Convolutional layer & 8 filters, $3\times3$ kernel size, $1\times1$ stride, ReLU activation, dropout\\
Flatten layer & 2048 neurons\\
Fully connected layer & 128 neurons, ReLU activation, dropout\\
Classification layer & 10 neurons, softmax activation\\
\hline
\end{tabular}
\caption{\textbf{Convolutional neural network architecture employed in this work.} The dropout rate for all layers is 7\,\%. The total number of parameters is 281\,818. 
ReLU is short for Rectified Linear Unit. 
}
\label{table:ai4stem_cnn}
\end{table*}

\clearpage

\beginsupplement

\section*{Automatic Identification of Crystal Structures and Interfaces via Artificial-Intelligence-based Electron Microscopy}

by Andreas Leitherer, Byung Chul Yeo, 
Christian H. Liebscher, and Luca M. Ghiringhelli

\section*{Supplementary Methods} 

To calculate the mismatch angle, the real-space lattices for both experimental images and reference image (taken from the training set) is reconstructed via atomap\cite{nord2017atomap}. 
The real-space lattice for the experimental data is extracted from the whole image, not for each local segment, which saves computation time. 
The local lattices that correspond to the local image patches from AI-STEM are extracted in a second step. 
For each of the local lattices, the rotation angle is determined that would align local and  reference lattices. 
To determine this angle,  point set registration is used. 
Specifically, the coherent point drift algorithm\cite{myronenko2010point} as implemented in the python package pycpd\cite{gatti2022pycpd}   is employed, where we use the rigid registration routine determining the amount of scaling, rotation and translation that is required to match two lattices. 
When comparing the lattices, to reduce the effects of boundary effects introduced by the fragmentation, we only use a small, spherical local region around the respective center (decreasing the radius until less than 20 atoms are contained, which corresponds to only few nearest neighbors, depending on the lattice symmetry). 
Depending on the initial relative lattice orientations, the algorithm may either perform a clockwise or counter-clockwise rotation to match the lattices, leading to jumps in the calculated angles and checkerboard-type heatmaps. 
To solve this, we determine the minimal rotation that is required to match two lattices: 
the lattice symmetry provides bounds for the maximum mismatch angle. 
For instance, hcp has  a 60 degrees rotational symmetry and a reference lattice can be rotated by a maximum of 30 degrees to match the hexagonal lattice. 
If the rotation angle determined via point set registration is larger than 30 degrees, we subtract 30 degrees and take the absolute value. 
The angle calculated in this fashion corresponds to the minimum amount of rotation to match two lattices. 
These values are reported in Fig. \ref{fig:fig4}d, h, l. 
A Jupyter notebook for running this calculation is provided (cf. section ``Data availability''.

 \begin{figure*}[ht]
 \centering
 \includegraphics[width=\textwidth]{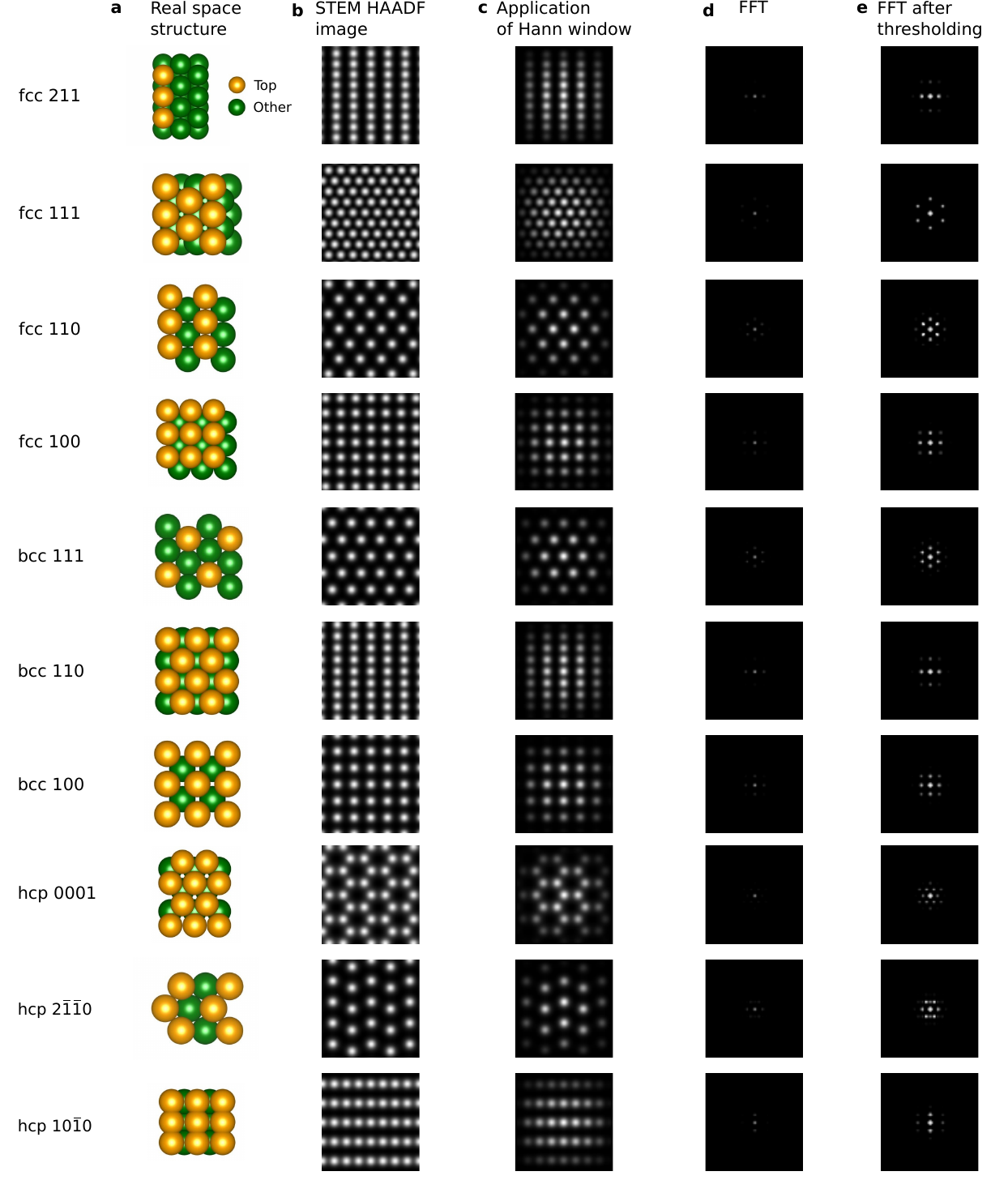}
 \caption{
\textbf{FFT-HAADF descriptor calculation for all 10 crystalline surfaces used in this work.} 
Given the real-space, 3D atomic structures (\textbf{a}), STEM HAADF images are simulated (\textbf{b}). Then,  a Hann window is applied (to reduce boundary artifacts, \textbf{c}), after which the FFT is calculated (\textbf{d}). Finally, a thresholding procedure is applied to enhance the peaks surrounding the dominant central, low-frequency contributions. In the figures of atomic structures (\textbf{a}), orange and green circles indicate atoms that belong to top and bottom layers, respectively. All HAADF image sizes are $1.2\times1.2\,\text{nm}^2$ (100 pixels). For the lattice parameters, we employ an interval of size 0.1\,\AA around the experimental values, where here, we show the images for the center of this interval:  for all Cu fcc single crystals, the lattice constant $a$ is 3.63 Å; for all Fe bcc single crystals, the lattice constant $a$ is 2.87 Å; for all Ti hcp single crystals, the lattice constants $a$ and $c$ are 2.95 Å and 4.68 Å, respectively ($c/a\sim 1.587$).
 }
 \label{fig:figS1}
 \end{figure*}

  \begin{figure*}[ht]
 \centering
 \includegraphics[width=\textwidth]{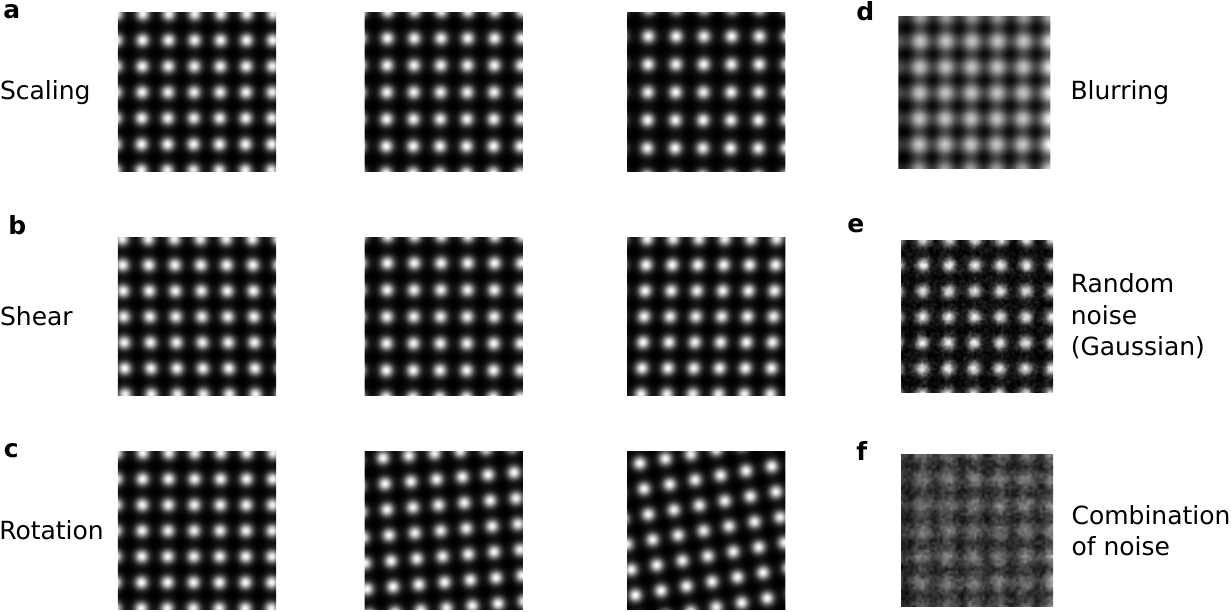}
 \caption{
\textbf{Visualization of data augmentation procedure.}
Considering the example of Fe bcc [100] (lattice parameter 2.87\,\AA), different augmentation steps are shown: scaling (\textbf{a}), shear (\textbf{b}),
rotations (\textbf{c}), blurring (\textbf{d}),
random noise (\textbf{e}, Gaussian), 
and a combination of all (\textbf{f}).
  \label{fig:figS2}
 }
 \end{figure*}

   \begin{figure*}[ht]
 \centering
 \includegraphics[width=\textwidth]{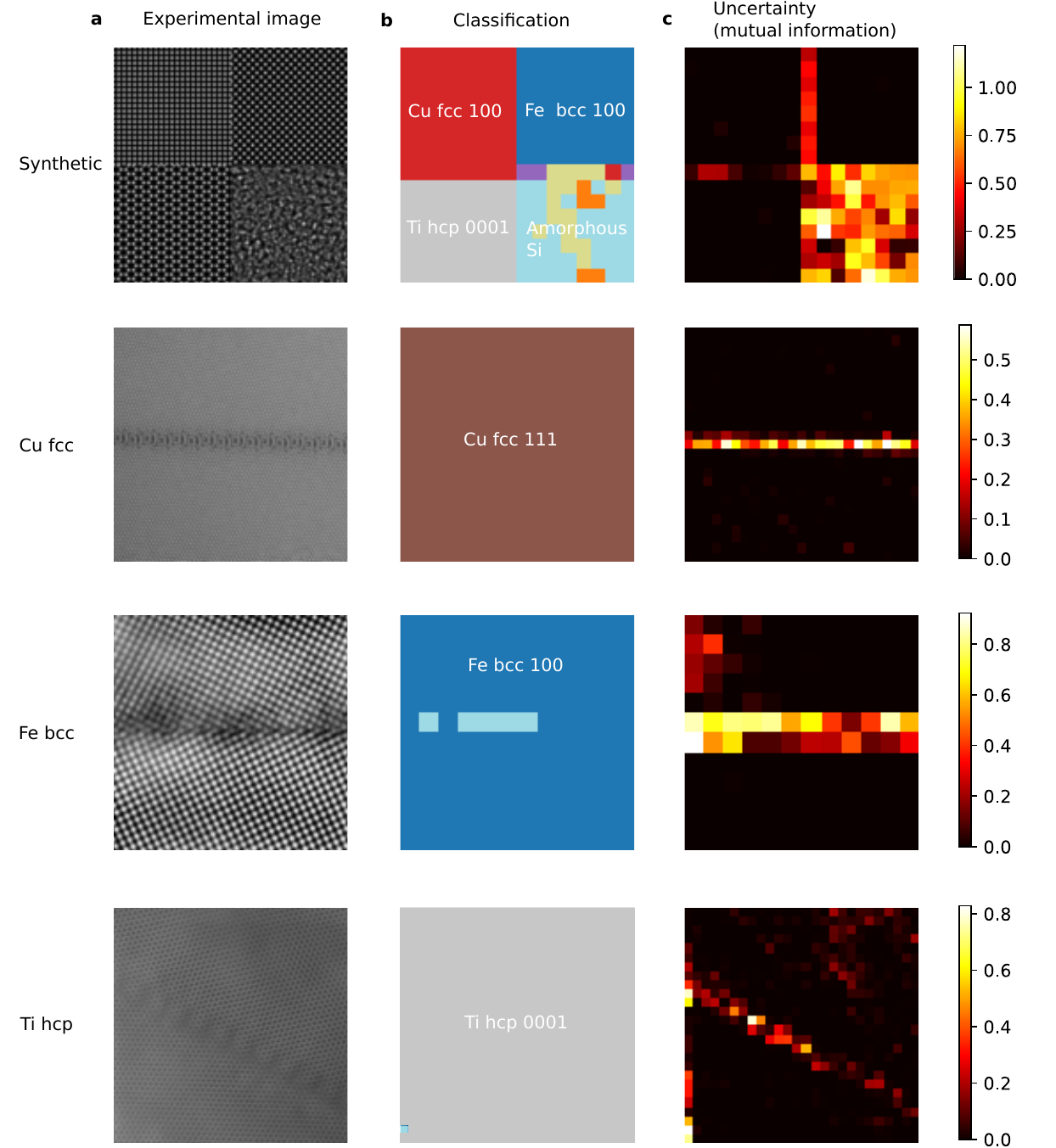}
 \caption{
\textbf{Effect of stride on AI-STEM's resolution.}
For the synthetic and experimental examples considered in Fig. \ref{fig:fig4} in the main text, we increase the stride by a factor of three (from 12 pixels to 36). This reduces the resolution of the interface regions but still allows to detect the main characteristics (i.e., bulk versus interface regions and correct assignment of lattice symmetry and orientation).
  \label{fig:figS3}
 }
 \end{figure*}

   \begin{figure*}[ht]
 \centering
 \includegraphics[width=\textwidth]{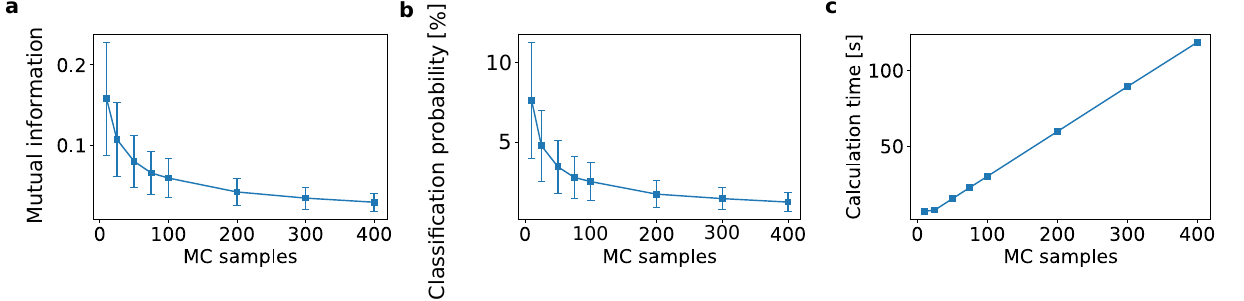}
 \caption{
\textbf{Choosing the number of Monte Carlo (MC) samples.}
When calculating classification probabilities (cf. Eq. \ref{eq:3}) and uncertainty estimates such as the mutual information (cf. Eq. \ref{equation:mutual_information}), 
the number of MC samples $T$ needs to be chosen. 
In particular, a value has to be identified, for which the above quantities show convergent behavior with respect to $T$. 
 As test bed, we employ the amorphous image (cf. Fig. \ref{fig:fig3}a), where we expect that the model is maximally uncertain. 
 This is an ideal scenario since differences between MC samples are expected to be maximal and thus a large $T$ may be required for convergence.   
 We investigate for each of the in total 576 local regions (out of the 2304 local regions of the whole image) a range of Monte Carlo samples (10, 25, 50, 75, 100, 200, 300, 400). 
 For each of the MC samples,  we calculate classification probabilities and mutual information for 5 iterations.  
We obtain 10 classification probabilities and focus on the maximal probability, which is most important for correct classification. 
 For each local window, we thus have 5 values for mutual information and (maximal) classification probability. 
 To quantify the spread over the iterations, we calculate the standard deviation, resulting in 576 values. 
Then we calculate the mean and standard deviation of all local-window standard deviations.
This way, we obtain for each MC sample two numbers that quantify the convergence.
In \textbf{a} the above-described calculation is shown for the mutual information, where the points correspond to the mean and the error bars correspond to the standard deviation of the local-window standard deviations.
Similarly, the maximal classification probability is shown in \textbf{b}. We observe onset of convergence (i.e., decreasing mean and standard deviation) around values of 100, which is the value we employ for the reported results. 
In particular, for $T=100$, the mutual-information mean and standard deviation is below the threshold of 0.1 that is employed for distinguishing interface and boundary regions (cf. Fig. \ref{fig:fig4}). 
Increasing this value provides only small improvement while increasing computational cost (that scales linearly with the number of MC samples $T$, cf. \textbf{c}). 
For instance, for $T=100$, one calculation (classification probabilities and mutual information for the whole image, i.e., 2304 local windows) takes around $\sim29.92$ seconds and doubling the number of MC samples doubles also the computation time to $59.52$ seconds. 
Note, however, that computation time is in principle not an issue since the calculation of predictions is trivial to parallelize. For the calculations we employed 
1 GPU (Tesla Volta V100 32GB) on the Talos machine learning cluster (at  MPCDF).
  \label{fig:figS4}
 }
 \end{figure*}

  \begin{figure*}[ht]
 \centering
 \includegraphics[width=\textwidth]{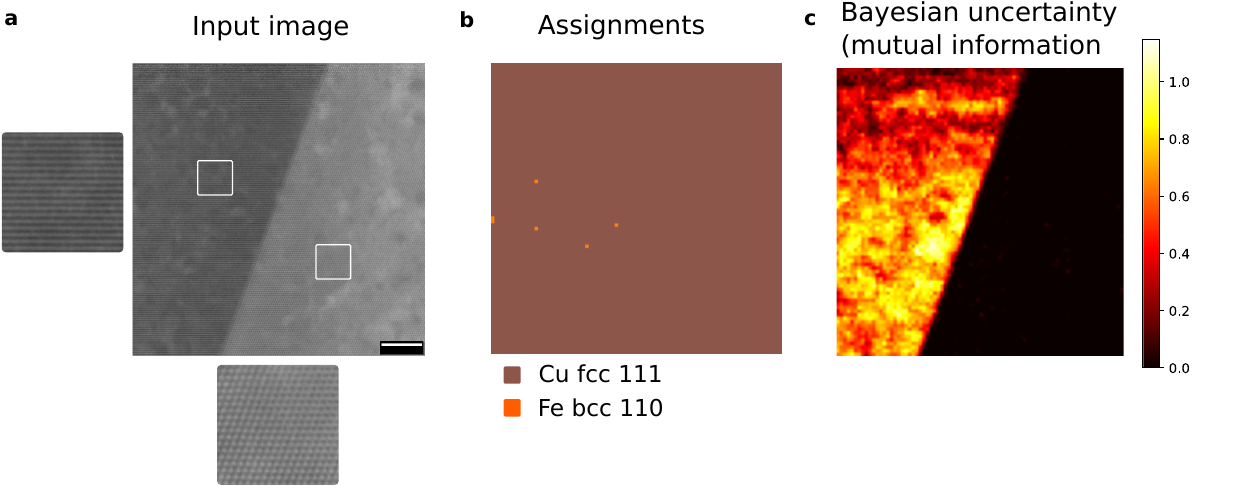}
 \caption{
\textbf{Effect of large deviations in crystal tilt on AI-STEM's performance.}
 \textbf{a} HAADF-STEM image of a grain boundary in aluminum imaged close to the [111] zone
axis orientation containing two grains. The left crystal is out off-zone axis only showing lattice fringes and the right one is
more closely aligned to the [111] zone axis. Two zoom-ins are shown (white windows and image patches on the left and bottom). The model predictions are shown in \textbf{b} (model assignments) and \textbf{c} (model uncertainty). For the grain with deviation from the perfect [111] orientation showing only lattice planes (\textbf{a}, left grain) the uncertainty is high, while the correct label can still be inferred from the entire image (with few exceptions). 
The comparatively high uncertainty indicates a low reliability in the classification requiring a careful interpretation of the prediction, which is related to the reduced projected crystal symmetry.  
A stride of 0.42\,nm (12 pixels) is used as well as a window size of 2.25\,nm (64 pixels, since the standard 1.2\,nm setting would result in a window size smaller than 64 pixels which is the minimum required size, determined by the FFT-HAADF descriptor; this window size is chosen since it is the closest to the values used in the main text results, enabling a better comparison). 
The scale bar in \textbf{a} is 5 nm. 
  \label{fig:figS5}
 }
 \end{figure*}

  \begin{figure*}[ht]
 \centering
 \includegraphics[width=\textwidth]{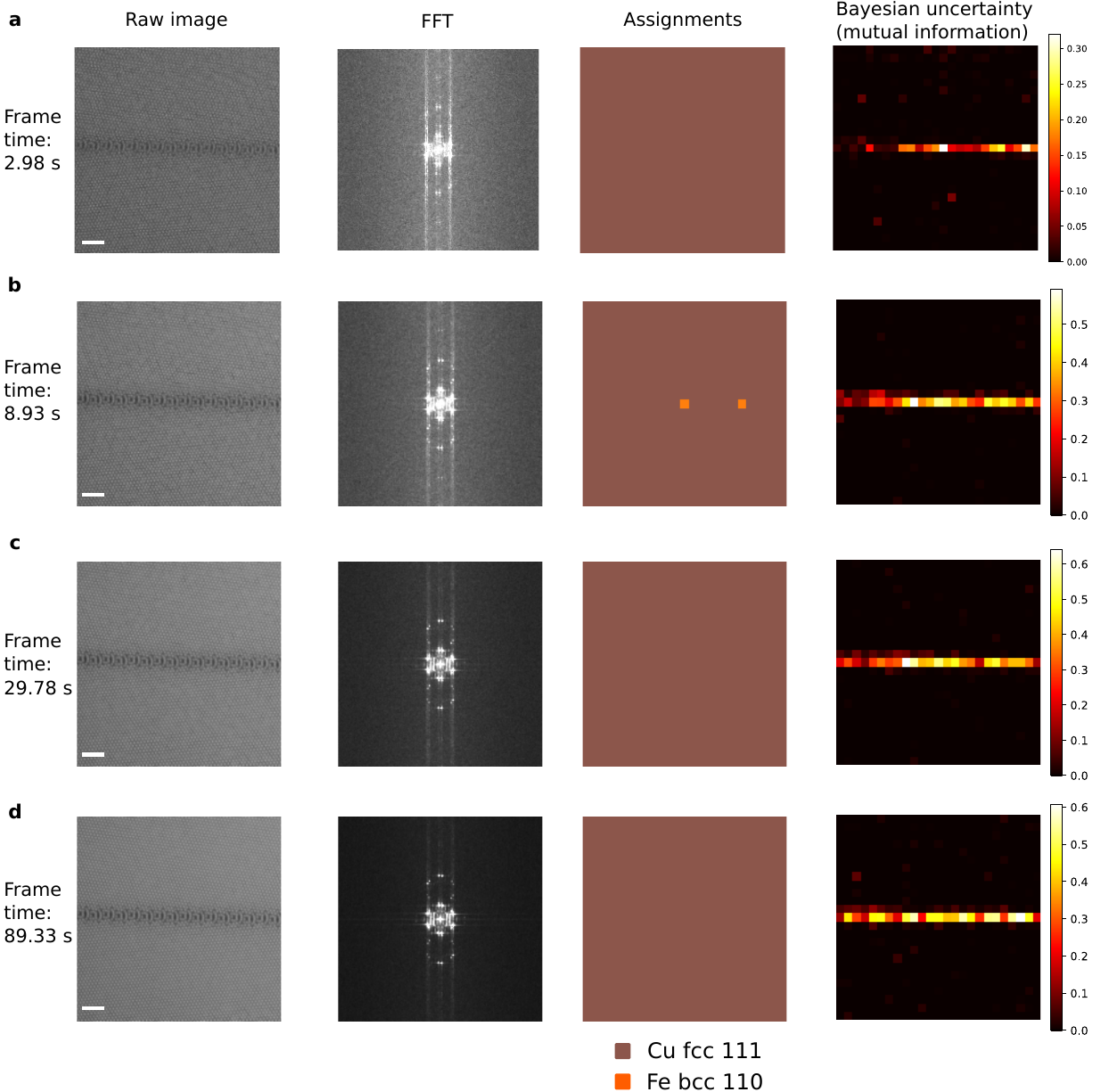}
 \caption{
\textbf{Effect of different levels of fast scan noise on AI-STEM's performance.}
Each row corresponds to a measurement for a different number of frames being averaged, where each column shows the raw real-space image, the FFT of the entire image, the neural-network assignments, as well as the mutual information. 
\textbf{a} shows a single image without frame averaging, while in  \textbf{b} 3 frames, \textbf{c} 10 frames, and in \textbf{d} 30 frames are averaged (corresponding to the final image in Fig. \ref{fig:fig4}a). The images contain different levels of fast scan noise and are thus a good test for the performance of AI-STEM with varying imaging conditions and noise contributions. For instance, the vertical bands in the FFTs represent different noise contributions primarily pertaining to fast scan noise. With an  increased number of frame averages, this noise level is reduced. AI-STEM's performance is essentially unaffected, with only small differences at the interface (or within the grains in \textbf{a}). The scale bar is 1\,nm in all raw real-space images.
  \label{fig:figS6}
 }
 \end{figure*}

\end{document}